%% file: ms.tex
\documentclass[a4paper,twoside,runningheads]{llncs}

\newif\ifarxiv
\arxivtrue

\usepackage[T1]{fontenc}
\usepackage{times}
\ifarxiv
\newcommand{\draftonly}[1]{}
\else
\newcommand{\draftonly}[1]{#1}
\fi
\usepackage{hyperref}
\usepackage{amsmath}
\usepackage{amssymb}
\usepackage{amsthm}
\usepackage{stmaryrd}
\usepackage{laws}

\usepackage{imakeidx}
\makeindex

\makeatletter
\newcounter{colwidth}
\newenvironment{eqncolumns}{\@ifnextchar[{\@eqncolumns}{\@@eqncolumns}}{\end{eqnarray}\end{minipage}\vspace{1ex}}
\def\@eqncolumns[#1]{\setcounter{colwidth}{100-#1}\mbox{}\vspace{-4ex}\\\begin{minipage}[t]{0.#1\textwidth}\begin{eqnarray}}
\def\@@eqncolumns{\setcounter{colwidth}{5}\mbox{}\vspace{-4ex}\\\begin{minipage}[t]{0.5\textwidth}\begin{eqnarray}}
\newcommand{\secondcolumn}{\end{eqnarray}\end{minipage}\begin{minipage}[t]{0.\thecolwidth\textwidth}\begin{eqnarray}}
\makeatother

\newcommand{\nondet}{\mathbin{\vee}}
\newcommand{\Nondet}{\mathop{\bigvee}}
\newcommand{\Seq}{\mathbin{;}}
\newcommand{\together}{\mathbin{\Cap}}
\newcommand{\meet}{\mathbin{\land}}

\newcommand{\sync}{\mathbin{\otimes}}
\newcommand{\refsto}{\mathrel{\succeq}}

\newcommand{\SyncAtomicId}{\iota}
\newcommand{\SyncId}{\Om{\SyncAtomicId}}

\newcommand{\kw}[1]{\mathsf{#1}}
\newcommand{\Nil}{\boldsymbol{\tau}}
\newcommand{\Atom}[1]{\mathsf{#1}}
\newcommand{\Ata}{\Atom{a}}
\newcommand{\Atb}{\Atom{b}}
\newcommand{\Patx}{\Atom{x}}
\newcommand{\Abort}{\lightning}
\newcommand{\Magic}{\kw{magic}}
\newcommand{\Term}{\kw{term}}
\newcommand{\Fair}{\kw{fair}}
\newcommand{\Idle}{\kw{idle}}
\newcommand{\Pre}[1]{\{#1\}}

\newcommand{\Rely}{\mathop{\kw{rely}}}
\newcommand{\rely}[1]{\Rely #1}
\newcommand{\Guar}{\mathop{\kw{guar}_{\boldsymbol{\pi}}}}
\newcommand{\guar}[1]{\Guar #1}
\newcommand{\Evolve}{\mathop{\kw{evolve}}}
\newcommand{\evolve}[1]{\Evolve #1}
\newcommand{\inv}[1]{\mathop{\kw{inv}} #1}

\newcommand{\Fin}[1]{#1^{\star}}
\newcommand{\Inf}[1]{#1^{\infty}}
\newcommand{\Om}[1]{#1^{\omega}}

\newcommand{\nat}{\mathbb{N}}

\newcommand{\cgd}[1]{\mathop{\tau} #1}
\newcommand{\estepd}{\epsilon}
\newcommand{\pstepd}{\pi}
\newcommand{\stepd}{\alpha}
\newcommand{\cstepd}{\boldsymbol{\stepd}}
\newcommand{\cpstepd}{\boldsymbol{\pstepd}}
\newcommand{\cestepd}{\boldsymbol{\estepd}}
\newcommand{\cpstep}[1]{\mathop{\pstepd}#1}
\newcommand{\cestep}[1]{\mathop{\estepd}#1}
\newcommand{\cstep}[1]{\mathop{\stepd}#1}
\newcommand{\universalrel}{\mathsf{univ}}
\newcommand{\prer}[1]{{}^\backprime#1}
\newcommand{\postr}[1]{#1'}
\newcommand{\Id}{\mathsf{id}}
\newcommand{\defs}{\mathrel{\widehat{=}}}

\newcommand{\spot}{\mathbin{.}}
\makeatletter
\def\comp@sym{\raise 0.6ex\hbox{\small\oalign{\hfil%
        $\scriptscriptstyle\mathrm{o}$\hfil%
        \cr\hfil$\scriptscriptstyle\mathrm{9}$\hfil}}}
\newcommand{\semi}{\mathrel{\comp@sym}}
\makeatother

\newcommand{\inter}{\mathbin{\cap}}
\newcommand{\union}{\mathbin{\cup}}

\makeatletter
\def\Spec{\@ifnextchar*{\@Spec}{\@@Spec}}
\def\@Spec*#1#2#3{\ifx\@empty#1\else#1\colon\fi
   [{#2}\ifx\@empty#2\else,~\fi#3]}
\def\@@Spec#1#2#3{\ifx\@empty#1\else
   \begin{array}{@{}l@{}}#1\colon\end{array}\!\!\fi%
   \left[{\begin{array}{@{}l@{}}#2\end{array}}\ifx\@empty#2\else,~\fi
   \begin{array}{@{}l@{}}#3\end{array}\right]}
\makeatother

\makeatletter
\newcommand{\ChainRel}[1]{\crcr 
  #1~ &
  \@ifnextchar*{\@ChainRelCommment}{}}
\newcommand{\Why}[1]{\mbox{{\color{blue}\hspace*{1em}#1}}}
\def\@ChainRelCommment*[#1]{\Why{#1}
  \crcr & 
  }
\newcommand{\StartRef}[1]{\hspace*{-1.5em}(\ref{#1}) \refsto
  \@ifnextchar[{\@StartRefCommment}{}}
\def\@StartRefCommment[#1]{\mbox{#1}
  \crcr 
  }
\makeatother

\newcommand{\Equals}{\ChainRel{=}}

\makeatletter
\def\@setmcodes#1#2#3{{\count0=#1 \count1=#3
  \loop \global\mathcode\count0=\count1 \ifnum \count0<#2
  \advance\count0 by1 \advance\count1 by1 \repeat}}

\DeclareSymbolFont{italic}{OT1}{\rmdefault}{m}{it}
\let\mathit\undefined
\DeclareSymbolFontAlphabet{\mathit}{italic}
\edef\@tempa{\hexnumber@\symitalic}
\@setmcodes{`A}{`Z}{"7\@tempa41}
\@setmcodes{`a}{`z}{"7\@tempa61}
\makeatother

\usepackage[colorinlistoftodos,textwidth=2cm]{todonotes}

\definecolor{Aqua}{rgb}{0,1,1}

\newcounter{hours}
\newcounter{minutes}
\newcommand{\printtime}{%
  \ifthenelse{\value{hours}<10}{0}{}\thehours:%
  \ifthenelse{\value{minutes}<10}{0}{}\theminutes}

\title{Reasoning about distributive laws \\ in a concurrent refinement algebra}
\author{
Larissa A. Meinicke\orcidID{0000-0002-5272-820X}
\and
Ian J. Hayes\orcidID{0000-0003-3649-392X}
}

\newbox{\MyDate}
\savebox{\MyDate}{\draftonly{ (\today\ \printtime)}}
\titlerunning{Distributive laws for concurrent refinement \usebox{\MyDate}}
\authorrunning{L. A. Meinicke \and I. J. Hayes \usebox{\MyDate}}
\institute{
School of Electrical Engineering and Computer Science, \\ 
The University of Queensland, Brisbane, Queensland 4072, Australia
  \draftonly{\\\vspace*{2ex} \today~\printtime}
}

\begin{document}

\maketitle

\draftonly{\vspace{-3ex}}
\begin{abstract}
Distributive laws are important for algebraic reasoning in arithmetic and logic.
They are equally important for algebraic reasoning about concurrent programs.
In existing theories such as Concurrent Kleene Algebra,
only partial correctness is handled, and many of its distributive laws are weak, 
in the sense that they are only refinements in one direction, rather than equalities.
The focus of this paper is on strengthening our theory to support the proof of strong distributive laws that are equalities,
and in doing so come up with laws that are quite general.
Our concurrent refinement algebra supports total correctness by allowing both finite and infinite behaviours.
It supports the rely/guarantee approach of Jones by encoding rely and guarantee conditions
as rely and guarantee commands.
The strong distributive laws may then be used to distribute rely and guarantee commands
over sequential compositions and into (and out of) iterations.
For handling data refinement of concurrent programs, strong distributive laws are essential.
\draftonly{{\color{purple}\textit{Note to reviewers.}
The index are not intended to be included in the final paper
but is included here as it may aid reviewing.}}
\end{abstract}

\section{Introduction}

\paragraph{Rely/guarantee concurrency.}

The concurrent refinement algebra is intended to support the rely/guarantee style of reasoning of Jones \cite{Jones81d,jones83a,jones83b}.
To provide a compositional approach to reasoning about concurrent programs,
Jones makes use of a rely condition, $r$, a binary relation between states, 
that corresponds to an assumption that any interference on a thread from its environment, satisfies $r$.
Complementing this, each thread has a guarantee condition $g$, also a binary relation between states,
and the thread must ensure every program transition it makes satisfies $g$.
\reffig{rely-guar} gives an execution trace that satisfies a rely/guarantee specification.
For a set of parallel threads, the guarantee of each thread must imply the rely condition of every other thread.

\begin{figure}
\begin{center}
\input{rely-guar}
\caption{An execution trace (of a program in an environment) consisting of states $\sigma_0$ through $\sigma_7$. 
Transitions of the program are labeled $\pstepd$ and transitions of its environment are labeled $\estepd$.
A trace satisfies a specification with pre-condition $p$, post-condition $q$, rely condition $r$ and guarantee condition $g$,
if whenever the initial state satisfies $p$ and all environment transitions satisfy $r$, 
the post-condition $q$ (a binary relation between states) holds between the initial and final states, 
and every program transition satisfies $g$.
Importantly, a trace of a thread includes transitions of both itself ($\pstepd$) and its environment ($\estepd$) \cite{Aczel83}.
}\labelfig{rely-guar}
\end{center}
\end{figure}
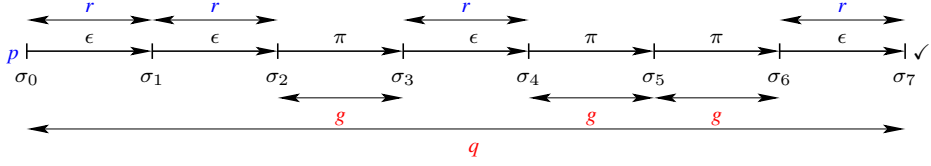

\paragraph{Refinement.}

The sequential refinement calculus \cite{BackWright98,Morgan94} 
encodes Floyd/Hoare preconditions and postconditions \cite{Floyd67,Hoare69a} via 
an assertion command, $\Pre{p}$, and a specification command, $\Spec{}{}{q}$, respectively.
It uses a refinement relation,%
\footnote{In the program algebra literature \cite{DBLP:journals/jlp/HoareMSW11} refinement is written, $c_1 \refsto c_2$,
but in the refinement calculus literature it is written, $c_1 \sqsubseteq c_2$. We follow the former convention in this paper.}
$c_1 \refsto c_2$,
meaning $c_1$ is refined (or implemented by) $c_2$.
The Hoare triple,%
\footnote{Note that the use of braces in a Hoare triple differs from their use for an assertion command.
This is the only place where we use a Hoare triple.}
$p \{ c \} q$, is encoded as the refinement, $\Pre{p} \Seq \Spec{}{}{q} \refsto c$.

For rely and guarantee conditions we follow the same approach as for preconditions and postconditions,
and encode them as the commands, $\rely{r}$ and $\guar{g}$, respectively.
A guarantee command, $\guar{g}$, ensures every program ($\pstepd$) transition from state $\sigma$ to state $\sigma'$ 
satisfies the binary relation between states $g$, that is, $(\sigma,\sigma') \in g$.
It puts no constraints on environment ($\estepd$) transitions.
A rely command, $\rely{r}$, represents an assumption that every environment ($\estepd$) transition
from $\sigma$ to $\sigma'$ satisfies the binary relation $r$, that is $(\sigma,\sigma') \in r$.
In the sequential refinement calculus, an assertion command, $\Pre{p}$,
either acts as a no-operation if $p$ holds or irrecoverably aborts if $p$ does not hold,
where aborting behaviour corresponds to Dijkstra's abort command \cite{Dijkstra75,Dijkstra76}.
The same approach is used for a rely command: 
it allows any transitions unless the environment makes a transition not satisfying $r$, 
in which case the rely command then aborts.

\paragraph{Weak conjunction.}

The satisfaction of a rely/guarantee specification 
with precondition $p$, postcondition $q$, guarantee condition $g$, and rely condition $r$ by a command, $c$, 
is written using the weak conjunction operator, $\together$, as the refinement,
\begin{align}
  \rely{r} \together \guar{g} \together \Pre{p} \Seq \Spec{}{}{q} \refsto c . \labelprop{rely-guar-sat}
\end{align}
where each transition of the weak conjunction of two commands, $c_1 \together c_2$, 
must be a transition of both $c_1$ and $c_2$, unless either $c_1$ or $c_2$ aborts, 
in which case $c_1 \together c_2$ aborts.
For example, the command $\rely{r} \together c$ behaves as $c$ unless the environment makes a transition not satisfying $r$,
in which case the rely command aborts, and hence the whole command, $\rely{r} \together c$, irrecoverably aborts, 
so that any behaviour whatsoever is possible from that point on.

Commands form a lattice for which the lattice join, $c_1 \nondet c_2$, is (non-deterministic) choice, that can behave as either $c_1$ or $c_2$,
and the lattice meet, $c_1 \meet c_2$, is a strong form of conjunction that has the common behaviours of $c_1$ and $c_2$.
Dijkstra's abort command, written $\Abort$ here, allows any behaviour whatsoever 
and hence it is the top command in the lattice,
and hence $\Abort \meet c = c$ for any command $c$, so that $c_1 \meet c_2$ can only abort if both $c_1$ and $c_2$ agree to abort.
When combining a rely command, which may abort, with the remainder of a specification, 
one needs to ensure that if the rely command aborts the whole specification aborts
(i.e. when an assumption is not met, the remaining commitments are no longer obliged to be fulfilled).
This is achieved by using weak conjunction operator, $c_1 \together c_2$.
In particular, $\Abort \together c = \Abort$, as compared with strong conjunction where $\Abort \meet c = c$.
 If neither $c_1$ nor $c_2$ aborts,
every behaviour of $c_1 \together c_2$ is both a behaviour of $c_1$ and a behaviour of $c_2$,
that is weak and strong conjunction coincide for commands that never abort.
Weak conjunction is associative, commutative and idempotent.
To illustrate the difference between weak ($\together$) and strong ($\meet$) conjunction of commands,
consider combining two specifications
in the sequential refinement calculus \cite{Ward:SpecConj,Groves2002}.
\begin{align}
  \Pre{p_1} \Seq \Spec{}{}{q_1} \together \Pre{p_2} \Seq \Spec{}{}{q_2} & = \Pre{p_1 \inter p_2} \Seq \Spec{}{}{q_1 \inter q_2} \labelprop{spec-conj} \\
  \Pre{p_1} \Seq \Spec{}{}{q_1} \meet \Pre{p_2} \Seq \Spec{}{}{q_2} & = \Pre{p_1 \union p_2} \Seq \Spec{}{}{(p_1 \implies q_1) \inter (p_2 \implies q_2)} \labelprop{spec-sconj}  
\end{align}
For weak conjunction, if both preconditions $p_1$ and $p_2$ hold initially, both postconditions must hold on termination,
whereas for strong conjunction, if either precondition holds initially,
then if $p_1$ holds initially, $q_1$ must hold on termination,
and if $p_2$ holds initially, $q_2$ must hold on termination.
The property for strong conjunction follows from the fact that it is the lattice meet, 
and hence it must refine both specification commands.

\paragraph{Naming conventions and syntactic precedence of operators.}
Commands are represented by $c$ and $d$, 
sets of commands by $C$,
atomic commands by $\Ata$ and $\Atb$,
pseudo-atomic commands by $\Patx$,
sets of states by $p$,
and
binary relations between states by $g$, $r$ and $q$.
The above naming conventions also apply to subscripted forms of the above names.
We assume the binary operator $\nondet$ has the lowest precedence and 
sequential composition ($\Seq$) has the highest precedence 
but otherwise make no assumptions about operator precedence
and use parentheses to disambiguate.

\paragraph{Distributive laws for rely/guarantee concurrency.}

In mathematics, distributive laws are important in algebraic reasoning, 
for example, in arithmetic, 
  $x*(y+z) = x*y + x*z$, 
and in logic, 
  $p \land (q \lor r) = (p \land q) \lor (p \land r)$.
Distributive laws hold an equally important place in algebraic reasoning about programs \cite{HoareHayesEtcFull87},
for example, 
\begin{align}
  d \parallel (c_1 \nondet c_2) = d \parallel c_1 \nondet d \parallel c_2,  \labelprop{par-distrib-nondet}
\end{align}
where $\parallel$ is parallel composition and $\nondet$ is non-deterministic choice.

Both parallel composition and weak conjunction of a guarantee command 
distribute over sequential composition (\refprop*{guar-par-distrib-seq}--\refprop*{guar-conj-distrib-seq}),
as does a rely command (\refprop*{rely-par-distrib-seq}--\refprop*{rely-conj-distrib-seq}).
\begin{align}
  \guar{g} \parallel c_1 \Seq c_2 & = (\guar{g} \parallel c_1) \Seq (\guar{g} \parallel c_2) \labelprop{guar-par-distrib-seq} \\
  \guar{g} \together c_1 \Seq c_2 & = (\guar{g} \together c_1) \Seq (\guar{g} \together c_2) \labelprop{guar-conj-distrib-seq} \\
  \rely{r} \parallel c_1 \Seq c_2 & = (\rely{r} \parallel c_1) \Seq (\rely{r} \parallel c_2) \labelprop{rely-par-distrib-seq} \\
  \rely{r} \together c_1 \Seq c_2 & = (\rely{r} \together c_1) \Seq (\rely{r} \together c_2) \labelprop{rely-conj-distrib-seq}
\end{align}

\paragraph{The abstract synchronisation operator.}

In our theory, parallel composition and weak conjunction are both synchronous operators \cite{Milner83,FM2016atomicSteps,FMJournalAtomicSteps},
meaning they synchronise by combining the first transitions of each operand to give a transition of their composition 
before synchronising the continuations of the two operand commands.
Parallel and weak conjunction share many axioms, 
so in our theory we introduce an abstract synchronisation operator, $\sync$, that has just their shared axioms.
For example, $\sync$ is associative and commutative and,
abstracting \refprop{par-distrib-nondet}, the abstract synchronisation operator distributes over choice,
\begin{align}
  d \sync (c_1 \nondet c_2) & = d \sync c_1 \nondet d \sync c_2 \labelprop{sync-distrib-nondet}
\end{align}
and hence $\sync$ is monotone in both arguments.
Any laws that are proven for $\sync$ can be applied to either parallel or weak conjunction.
For example, \refprop{guar-par-distrib-seq} and \refprop{guar-conj-distrib-seq} are both instances of \refprop{guar-sync-distrib-seq},
and \refprop{rely-par-distrib-seq} and \refprop{rely-conj-distrib-seq} are both instances of \refprop{rely-sync-distrib-seq}.
\begin{align}
  \guar{g} \sync c_1 \Seq c_2 & = (\guar{g} \sync c_1) \Seq (\guar{g} \sync c_2) \labelprop{guar-sync-distrib-seq} \\
  \rely{r} \sync c_1 \Seq c_2 & = (\rely{r} \sync c_1) \Seq (\rely{r} \sync c_2) \labelprop{rely-sync-distrib-seq} 
\end{align}
One of the challenges addressed in this paper is to devise suitable restrictions on a command $d$ to ensure,
\begin{align}
  d \sync c_1 \Seq c_2 & = (d \sync c_1) \Seq (d \sync c_2) \labelprop{d-sync-seq0}
\end{align}
so that \refprop{guar-sync-distrib-seq} and \refprop{rely-sync-distrib-seq} are instances of \refprop{d-sync-seq0}.

\paragraph{Weak distributive laws.}

It is straightforward to show the weak distributive law,
\begin{align}
  d \sync c_1 \Seq c_2 \refsto (d \sync c_1) \Seq (d \sync c_2) && \mbox{if } d \refsto d \Seq d \labelprop{weak-sync-distrib-seq}
\end{align}
because synchronisation satisfies a weak interchange axiom with sequential composition,%
\footnote{Both $\parallel$ and $\together$ satisfy this weak interchange axiom with sequential composition.} 
similar to that in Concurrent Kleene Algebra \cite{DBLP:journals/jlp/HoareMSW11}:
\begin{align}
  d_1 \Seq d_2 \sync c_1 \Seq c_2 \refsto (d_1 \sync c_1) \Seq (d_2 \sync c_2). \labelax{sync-interchange-seq}
\end{align}
Law \refprop{weak-sync-distrib-seq} follows from the assumption ($d \refsto d \Seq d$),
monotonicity of $\sync$, and \refax{sync-interchange-seq}:
\begin{align*}
  d \sync c_1 \Seq c_2 
 & \refsto
  d \Seq d \sync c_1 \Seq c_2 
  \refsto 
  (d \sync c_1) \Seq (d \sync c_2) .
\end{align*}
Fixed iteration, $c^i$, of a command $c$, $i$ times for $i \in \nat$, is defined inductively by,
\begin{eqncolumns}
  c^0 & = \Nil \labelax{iter-zero} 
\secondcolumn
  c^{i+1} & = c \Seq c^i \labelax{iter-succ}
\end{eqncolumns}
where $\Nil$ is the command that terminates immediately from any state;
it is the identity of sequential composition,
that is, $\Nil \Seq c = c = c \Seq \Nil$ for any command $c$.
Weakly distributing $d$ into a fixed iteration holds for all $i \in \nat$ as follows.
\begin{align}
  d \sync c^i & \refsto (d \sync c)^i & \mbox{if } d \sync \Nil \refsto \Nil \mbox{ and } d \refsto d \Seq d \labelprop{weak-sync-distrib-fixed-iter}
\end{align}
The proof is by induction on $i$:
the base case for $i$ zero is handled by the first assumption, $d \sync \Nil \refsto \Nil$, 
and
the inductive case uses \refax{iter-succ}, \refprop{weak-sync-distrib-seq} using the second assumption, the inductive hypothesis and finally \refax{iter-succ}:
\begin{align*}
  d \sync c^{i+1} 
  = d \sync c \Seq c^i 
  \refsto (d \sync c) \Seq (d \sync c^i) 
  \refsto (d \sync c) \Seq (d \sync c)^i
  = (d \sync c)^i .
\end{align*}

Finite iteration of a command $c$, zero or more times, corresponds to the choice over all the fixed iterations, $c^i$, for $i \in \nat$,
\begin{align}
  \Fin{c} = \Nondet_{i \in \nat} c^i . \labeldef{iter-finite}
\end{align}
Weakly distributing $d$ into a finite iteration holds under the same assumptions as for fixed iteration.
\begin{align}
  d \sync \Fin{c} \refsto \Fin{(d \sync c)}  && \mbox{if } d \sync \Nil \refsto \Nil \mbox{ and } d \refsto d \Seq d. \labelprop{weak-sync-distrib-finite-iter}
\end{align}
The proof decomposes $\Fin{c}$ using \refdef{iter-finite}, 
distributes the synchronisation into the choice,
applies \refprop{weak-sync-distrib-fixed-iter} for each $i \in \nat$, 
and then combines the results using \refdef{iter-finite}.
\begin{align}
  d \sync \Fin{c} 
  = d \sync \Nondet_{i \in \nat} c^i
  = \Nondet_{i \in \nat} (d \sync c^i)
  \refsto \Nondet_{i \in \nat} (d \sync c)^i
  = \Fin{(d \sync c)} .
\end{align}

\paragraph{Strong distributive laws for finite sequential compositions.}

While \refprop{weak-sync-distrib-seq}, \refprop{weak-sync-distrib-fixed-iter} and \refprop{weak-sync-distrib-finite-iter} are refinements,
there are situations where one needs the following stronger distributive laws that are equalities rather than refinements.
\begin{align}
  d \sync c_1 \Seq c_2 & = (d \sync c_1) \Seq (d \sync c_2) \labelprop{sync-distrib-seq} \\
  d \sync c^i & = (d \sync c)^i \labelprop{sync-distrib-fixed-iter} \\
  d \sync \Fin{c} & = \Fin{(d \sync c)} \labelprop{sync-distrib-finite-iter}
\end{align}
The main contribution of this paper is to investigate the restrictions on $d$ 
that allow these laws 
(and other distributive laws detailed below)
to be strengthened to equalities.
If both
\begin{align}
  d \sync \Nil & = \Nil \labelprop{d-sync-nil1} \\
  d \sync c_1 \Seq c_2 & = (d \sync c_1) \Seq (d \sync c_2) & \mbox{for any commands $c_1$ and $c_2$} \labelprop{d-sync-seq1}
\end{align}
then it is straightforward to show both 
\refprop{sync-distrib-fixed-iter} and \refprop{sync-distrib-finite-iter} hold
using proofs similar in structure to the proofs of \refprop{weak-sync-distrib-fixed-iter} and \refprop{weak-sync-distrib-finite-iter}
--- see \refsect{distrib-finite}.

\paragraph{Atomic commands.}

The concurrent refinement algebra \cite{FM2016atomicSteps,FMJournalAtomicSteps}
includes a subset of commands, know as \emph{atomic commands}, 
that can only perform a single transition and then terminate.
Possibly infinite iteration of a command $c$, zero or more times, is denoted $\Om{c}$;
it includes both finite and infinite iterations of $c$, where $\Inf{c}$ gives the infinite iteration of $c$.
\begin{align}
  \Om{c} = \Fin{c} \nondet \Inf{c}  \labelprop{isolation}
\end{align}
It turns out that if $d$ is either a finite iteration, $\Fin{\Ata}$, 
or a possibly infinite iteration, $\Om{\Ata}$, of an atomic command $\Ata$, then
\refprop{d-sync-nil1} and \refprop{d-sync-seq1} both hold, for example,
\begin{align}
  \Om{\Ata} \sync \Nil & = \Nil \labelprop{iter-sync-nil} \\
  \Om{\Ata} \sync c_1 \Seq c_2 & = (\Om{\Ata} \sync c_1) \Seq (\Om{\Ata} \sync c_2) & \mbox{for any commands $c_1$ and $c_2$.} \labelprop{iter-distrib-seq}
\end{align}
Hence \refprop{sync-distrib-fixed-iter} and \refprop{sync-distrib-finite-iter} both hold for $d$ as either $\Fin{\Ata}$ or $\Om{\Ata}$.
A program guarantee command, $\guar{g}$, can be defined as a possibly infinite iteration of an atomic command
that can perform a program transition from state $\sigma$ to state $\sigma'$ provided it satisfies $g$, (i.e.\ $(\sigma,\sigma') \in g$),
or any environment transition.
Therefore laws \refprop{sync-distrib-seq}, \refprop{sync-distrib-fixed-iter} and \refprop{sync-distrib-finite-iter}
hold for $d$ a guarantee command, $\guar{g}$ --- see \refsect{rely-guar}.

\paragraph{Pseudo-atomic commands.}

While an atomic command can make a single transition and terminate,
a \emph{pseudo-atomic command} can make a single transition and either terminate or abort.
Any pseudo-atomic command, $\Patx$, can be written in the form $\Ata \nondet \Atb \Seq \Abort$,
where $\Ata$ and $\Atb$ are atomic commands.
Such a command can make a single transition (of $\Ata$) and terminate
or make a single transition (of $\Atb$) and abort.
Pseudo-atomic commands share many properties with atomic commands.

It turns out that if $d$ is either a finite iteration, $\Fin{\Patx}$, of a pseudo-atomic command $\Patx$
or a possibly infinite iteration, $\Om{\Patx}$,  then
\refprop{d-sync-nil1} and \refprop{d-sync-seq1} both hold
and hence \refprop{sync-distrib-fixed-iter} and \refprop{sync-distrib-finite-iter} both hold.
A rely command, $\rely{r}$, is defined as a possibly infinite iteration of a pseudo-atomic command
that can make any program transition or any environment transition satisfying $r$ and terminate,
or any environment transition not satisfying $r$ and abort.
Therefore laws \refprop{sync-distrib-seq}, \refprop{sync-distrib-fixed-iter} and \refprop{sync-distrib-finite-iter}
hold for $d$ a rely command, $\rely{r}$ --- see \refsect{rely-guar}.

The everywhere infeasible command, $\Magic$, 
is considered to be a (degenerate) atomic command,
and $\Magic$ is a left annihilator for sequential composition,
that is, $\Magic \Seq c = \Magic$, for any command $c$,
and hence the atomic command $\Ata$ is 
a special case of a pseudo-atomic command because
$\Ata \nondet \Magic \Seq \Abort = \Ata \nondet \Magic = \Ata$,
as $\Magic$ is the identity of non-deterministic choice.
That is, atomic commands are a subset of pseudo-atomic commands.

\paragraph{Pseudo-atomic fixed points.}

Finite iteration, $\Fin{c}$, of a command $c$ is the least fixed point of the function,
\(
  (\lambda y \spot \Nil \nondet c \Seq y),
\)
and possibly infinite iteration, $\Om{c}$, is the greatest fixed point of the same function.
As a further generalisation, it turns out that if $d$ is any fixed point of the function,
\(
  (\lambda y \spot \Nil \nondet \Patx \Seq y), 
\)
where $\Patx$ is a pseudo-atomic command,
that is,
\begin{align}
  d = \Nil \nondet \Patx \Seq d,  \labeldef{pseudo-atomic-fixed-point0}
\end{align}
both \refprop{d-sync-nil1} and \refprop{d-sync-seq1} hold,
and hence \refprop{sync-distrib-fixed-iter} and \refprop{sync-distrib-finite-iter} also hold --- see \refsect{pseudo-atomic-fixed-points}.
We refer to commands, $d$, satisfying \refdef{pseudo-atomic-fixed-point0} as \emph{pseudo-atomic fixed points}.
Clearly, $\Fin{\Patx}$ and $\Om{\Patx}$ are both pseudo-atomic fixed points if $\Patx$ is a pseudo-atomic command,
and hence so are the special case forms $\Fin{\Ata}$ and $\Om{\Ata}$, for an atomic command $\Ata$.
The commands 
$\Term$ representing the most general command that can perform only a finite number of program transitions but does not constrain its environment, and
$\Fair$ representing the most general command that disallows preemption by the environment forever,
are both pseudo-atomic fixed points and hence both satisfy \refprop{d-sync-nil1} and \refprop{d-sync-seq1},
and hence \refprop{sync-distrib-fixed-iter} and \refprop{sync-distrib-finite-iter} --- see \refsect{rely-guar}.

\paragraph{Overview.}

\refsect{CRA} gives details of the synchronous algebra.
\refsect{distrib-finite} gives general laws for distributing into a finite iteration.
\refsect{pseudo-atomic-fixed-points} tackles proving the (strong) distributive laws 
\refprop{sync-distrib-seq}, \refprop{sync-distrib-fixed-iter} and \refprop{sync-distrib-finite-iter}
for $d$ a pseudo-atomic fixed point,
with the obvious corollaries that all three laws also hold for 
$d$ of the form $\Fin{\Patx}$  or $\Om{\Patx}$ for $\Patx$ a pseudo-atomic command,
and hence for $d$ of the form $\Fin{\Ata}$ or $\Om{\Ata}$ for an atomic command $\Ata$.
\refsect{rely-guar} applies these theorems to the rely/guarantee approach of Jones \cite{Jones81d,jones83a,jones83b}
as encoded in our concurrent refinement algebra.

\section{Synchronous algebra}\labelsect{CRA}

Our approach is based on a synchronous algebra \cite{Pris10,FM2016atomicSteps,FMJournalAtomicSteps},
which has been formalised in Isabelle/HOL \cite{IsabelleHOL}.%
\footnote{A trace model of the algebra \cite{DaSMfaWSLwC} has been developed in Isabelle/HOL 
and the axioms of our theory have been shown to be consistent with it.}
It distinguishes two subsets of commands:
\begin{itemize}
\item
instantaneous test commands, of the form $\cgd{p}$, that terminate immediately if the initial state is in the set of states $p$ but are infeasible otherwise;
these are similar to tests in Kozen's Kleene Algebra with Tests (KAT) \cite{kozen97kleene},
and
\item
atomic commands, denoted by $\Ata$ and $\Atb$, that may take a single transition and terminate;
transitions are labeled as either program ($\pstepd$) or environment ($\estepd$).
\end{itemize}
The everywhere infeasible command, $\Magic$, that has no behaviours at all,
corresponds to both the (least) test, $\cgd{\emptyset}$, that fails in every state;
$\Magic$ also corresponds to the atomic command that can make no transitions whatsoever.
The (greatest) test that succeeds terminating immediately in every state is the identity of sequential composition, $\Nil$.
An additional command is Dijkstra's abort command \cite{Dijkstra75,Dijkstra76}, denoted $\Abort$ here, that represents catastrophic failure;
after aborting no further assumptions can be made about the behaviour of the program ---
any behaviour is possible.
Abort is a left annihilator for sequential composition, 
i.e.\ $\Abort \Seq c = \Abort$, for any command $c$,
in particular, $\Abort \Seq \Magic = \Abort$, unlike some other theories \cite{DBLP:journals/jlp/HoareMSW11},
in which, $c \Seq \Magic = \Magic$, for any command $c$.
Abort is also an annihilator for $\sync$ and hence $\parallel$ and $\together$.

An assertion, $\Pre{p}$, acts as a no-operation unless its initial state does not satisfy $p$, in which case it aborts \refdef{assert},
where $\overline{p}$ denotes the set of states not in $p$.
A test absorbs a following assertion on the same set of states $p$ \refprop{test-assert}.
Tests and assertions satisfy a Galois connection
that allows an assertion on the left side of a refinement to be replaced by the corresponding test on the right side \refprop{galois-test-assert}.
\begin{align}
  \Pre{p} & \defs \Nil \nondet \cgd{\overline{p}} \Seq \Abort \labeldef{assert} \\
  \cgd{p} \Seq \Pre{p} & = \cgd{p} \labelprop{test-assert} \\
  \Pre{p} \Seq c \refsto d & \mbox{~~~if and only if~~~} c \refsto \cgd{p} \Seq d \labelprop{galois-test-assert}
\end{align}

The abstract synchronisation operator, $\sync$, is assumed to be associative and commutative, 
with atomic identity $\SyncAtomicId$ \refax{sync-atomic-id}, 
and (command) identity $\SyncId$ \refax{sync-id}.
Atomic steps are closed under synchronisation, as are tests,
where the synchronisation of two tests corresponds to their conjunction \refax{test-sync-test}.
A test distributes into a synchronisation \refax{test-distrib-sync}.
If synchronising $c_1$ with $\Nil$ is infeasible, 
so is synchronising any extension of $c_1$ \refax{whole-if-first}.
An atomic command cannot synchronise with $\Nil$ 
and hence their synchronisation corresponds to the every\-where infeasible command $\Magic$ \refax{nil-sync-atomic}.
The synchronisation of two commands that both start with atomic commands
first synchronises their respective atomic commands and then synchronises their continuations \refax{atomic-sync-atomic}.%
\footnote{Axiom \refax{atomic-sync-atomic} is the main axiom that signifies that $\sync$ is synchronous,
for example, $c \parallel d$ synchronises a program transition of $c$ with a matching environment transition of $d$, or vice versa, 
to give a program transition of $c \parallel d$,
and matching environment transitions of both $c$ and $d$ to give an environment transition of $c \parallel d$.
For a non-synchronous parallel operator there are no environment transitions and program transitions are interleaved.}
Synchronisation distributes over a non-deterministic choice over a non-empty set of commands \refax{sync-distrib-Nondet}.
Sequential composition distributes over a non-deterministic choice from the right \refax{seq-distrib-Nondet-right}
and from the left provided the set of commands is non-empty \refax{seq-distrib-Nondet-left}.
Synchronisation is abort strict \refax{sync-abort-strict}.
\begin{align}
  \SyncAtomicId \sync \Ata & = \Ata & \mbox{if $\Ata$ is an atomic command} \labelax{sync-atomic-id} \\
  \SyncId \sync c & = c \labelax{sync-id} \\
  \cgd{p_1} \sync \cgd{p_2} & = \cgd{(p_1 \inter p_2)}  \labelax{test-sync-test} \\
  \cgd{p} \Seq (c_1 \sync c_2) & = \cgd{p} \Seq c_1 \sync \cgd{p} \Seq c_2 \labelax{test-distrib-sync} \\
  \Nil \sync c_1 \Seq c_2 & = \Magic & \mbox{if } \Nil \sync c_1 = \Magic \labelax{whole-if-first} \\
  \Nil \sync \Ata & = \Magic \labelax{nil-sync-atomic} \\
  \Ata_1 \Seq c_1 \sync \Ata_2 \Seq c_2 & = (\Ata_1 \sync \Ata_2) \Seq (c_1 \sync c_2) \labelax{atomic-sync-atomic} \\
  (\Nondet_{c \in C} c) \sync d & = \Nondet_{c \in C} (c \sync d) & \mbox{if } C \neq \{\} \labelax{sync-distrib-Nondet} \\
  (\Nondet_{c \in C} c) \Seq d & = \Nondet_{c \in C} (c \Seq d) \labelax{seq-distrib-Nondet-right} \\
  d \Seq (\Nondet_{c \in C} c) & = \Nondet_{c \in C} (d \Seq c) & \mbox{if } C \neq \{\} \labelax{seq-distrib-Nondet-left} \\
  \Abort \sync c & = \Abort \labelax{sync-abort-strict}
\end{align}
From \refax{nil-sync-atomic} one can deduce \refprop{test-sync-atomic} because $\Nil \refsto \cgd{p}$, for any test $\cgd{p}$.
From \refax{sync-distrib-Nondet} with $C$ as $\{c_1,c_2\}$ one can deduce \refprop{sync-distrib-nondet},
and similarly \refprop{seq-distrib-nondet-right} from \refax{seq-distrib-Nondet-right}.
From \refax{sync-distrib-Nondet} one can deduce \refprop{sync-distrib-Nondet},
with its side condition handling the case when $C$ is empty.
A test on a branch of a synchronisation may be pulled out as a test on the whole,
if the other branch is not immediately aborting \refprop{test-command-sync-command}, i.e. if $c_1 \sync \Magic = \Magic$ \cite{FMJournalAtomicSteps}.
A test at the end of either branch of a synchronisation can be pulled out as a final test \refprop{final-test}.
An assertion at the start of either branch of a synchronisation can be pulled out of the synchronisation \refprop{assert-command-sync-command}.
\begin{align}
  \cgd{p} \sync \Ata & = \Magic  \labelprop{test-sync-atomic} \\
  (c_1 \nondet c_2) \Seq c & = (c_1 \Seq c) \nondet (c_2 \Seq c) \labelprop{seq-distrib-nondet-right} \\
  (\Nondet_{c \in C} c) \sync d & = \Nondet_{c \in C} (c \sync d) & \mbox{if } \Magic \sync d = \Magic \labelprop{sync-distrib-Nondet} \\
  c_1 \sync t \Seq c_2 & = t \Seq (c_1 \sync c_2) & \mbox{if } \Magic \sync c_1 = \Magic \labelprop{test-command-sync-command} \\
  c_1 \Seq t \sync c_2 & = (c_1 \sync c_2) \Seq t \labelprop{final-test} \\
  c_1 \sync \Pre{p} \Seq c_2 & = \Pre{p} \Seq (c_1 \sync c_2) \labelprop{assert-command-sync-command}
\end{align}

A pseudo-atomic command, $\Patx$, has laws similar to \refax{sync-atomic-id}, \refax{nil-sync-atomic} and \refax{atomic-sync-atomic}.
The following laws can be shown by expanding $\Patx$ to the form $(\Ata \nondet \Atb \Seq \Abort)$ for some atomic commands $\Ata$ and $\Atb$ 
and simplifying using the above axioms and laws.
\begin{align}
  \SyncAtomicId \sync \Patx & = \Patx \labelprop{pseudo-sync-atomic-id} \\ 
  \Nil \sync \Patx & = \Magic \labelprop{pseudo-sync-nil} \\
  \Patx_1 \Seq c_1 \sync \Patx_2 \Seq c_2 & = (\Patx_1 \sync \Patx_2) \Seq (c_1 \sync c_2) \labelprop{pseudo-sync-pseudo}
\end{align}

\paragraph{Expanded form.}

An important axiom is that the behaviour of any command, $c$, can be decomposed into,
\begin{itemize}
\item
an immediately aborting behaviour from initial states not in some set $p_n$,
\item
an immediately terminating behaviour from initial states in some set $p_t$, 
or
\item
$c$ performs a transition of some atomic command, $\Ata$, and then behaves as some continuation command, $c'$.
\end{itemize}
In the final case above, the continuation behaviour, $c'$, depends on which atomic command, $\Ata$, is taken,
and hence a choice over a set, $C$, of pairs of $\Ata$ and $c'$ is required.
In summary, for any command $c$, there exist sets of states $p_n$ and $p_t$ 
and a set $C$ of pairs of atomic command and command, such that,%
\footnote{For readers familiar with labeled transition systems, 
$\overline{p_n}$ represents the set of states in which $c$ aborts,
$p_t$ the set of states in which $c$ can terminate,
and for an pair of commands $(\Ata,c')$ in $C$, 
$c$ can do a transition of $\Ata$ and then behave as $c'$.
The main difference is that a pair $(\Ata,c')$ groups together transitions (of $\Ata$) that have the same continuation, $c'$.
We also include the special command $\Abort$, which is not usually available in labelled transition systems
but is necessary to represent failure of assertions and rely commands.
}
\begin{align}
  c = \Pre{p_n} \Seq(\cgd{p_t} \nondet \Nondet_{(\Ata,c') \in C} (\Ata \Seq c'))
  \labelax{expanded-form}
\end{align}
Note that aborting allows any possible behaviour and thus subsumes the other two alternatives.
The set $p_t$ can be $\emptyset$, if no immediately terminating behaviours are possible for $c$,
$p_a$ can be the set of all states, if no immediately aborting behaviours are possible for $c$,
and
the set of pairs $C$ can be empty if $c$ cannot make any transitions.

Straightforward calculations give that, if $c$ corresponds to the expanded form in \refax{expanded-form}, the following laws hold.
\begin{align}
\Nil \sync c
& = \Pre{p_n} \Seq \cgd{p_t}  \labelprop{expanded-nil} \\
\Ata_1^{k+1} \sync c
& = \Pre{p_n} \Seq \Nondet_{(\Ata,c') \in C} ((\Ata_1 \sync \Ata) \Seq (\Ata_1^k \sync c'))
& \mbox{for } k \in \nat  \labelprop{expanded-succ}
\end{align}

\section{Distributing into finite iteration}\labelsect{distrib-finite}

In this section we show that, 
if a command $d$ satisfies the following two properties,
\begin{align}
  d \sync \Nil & = \Nil  \labelprop{d-sync-nil} \\
  d \sync c_1 \Seq c_2 & = (d \sync c_1) \Seq (d \sync c_2) \labelprop{d-sync-seq}
\end{align}
it distributes into fixed and finite iteration.
\refsect{pseudo-atomic-fixed-points} then focuses on properties of $d$ that ensure it satisfies the above two properties.
\begin{theoremx}[distrib-fixed-iter]
For all $i \in \nat$,
\begin{align}
  d \sync c^0 & = (d \sync c)^0 & \mbox{if } d \sync \Nil = \Nil \labelprop{sync-iter-zero} \\
  d \sync c^{i+1} & = (d \sync c)^{i+1} & \mbox{if } \forall c_1, c_2 \spot d \sync c_1 \Seq c_2 = (d \sync c_1) \Seq (d \sync c_2) \labelprop{sync-iter-succ}   
\end{align}
\end{theoremx}

\begin{proof}
For \refprop{sync-iter-zero}, using \refax{iter-zero} and the assumption,
$d \sync c^0 = d \sync \Nil = \Nil = (d \sync c)^0$.
Property \refprop{sync-iter-succ} is shown by induction on $i$.
It holds trivially for $i=0$. 
Assuming the property for $i$, we show it holds for $i+1$ 
using \refax{iter-succ}, the assumption, the induction hypothesis and finally \refax{iter-succ}:
$d \sync c^{i+2}
= d \sync c \Seq c^{i+1}
= (d \sync c) \Seq (d \sync c^{i+1})
= (d \sync c) \Seq (d \sync c)^{i+1}
= (d \sync c)^{i+2}
$.
\end{proof}
The same two assumptions also ensure $d$ distributes into a finite iteration.
\begin{theoremx}[distrib-finite-iter]
If $d \sync \Nil = \Nil$ and $d \sync c_1 \Seq c_2 = (d \sync c_1) \Seq (d \sync c_2)$ for any commands $c_1$ and $c_2$ then,
\(
  d \sync \Fin{c} = \Fin{(d \sync c)} .
\)
\end{theoremx}

\begin{proof}
The proof decomposes $\Fin{c}$ using \refdef{iter-finite}, 
distributes over the choice using \refax{sync-distrib-Nondet},
applies \Theorem*{distrib-fixed-iter} for each $i$ using the assumptions,
and
composes the results using \refdef{iter-finite}:
$d \sync \Fin{c} 
= d \sync \Nondet_{i \in \nat} c^i 
= \Nondet_{i \in \nat} (d \sync c^i) 
= \Nondet_{i \in \nat} (d \sync c)^i 
= \Fin{(d \sync c)}
$.
\end{proof}

\section{Pseudo-atomic fixed point distributive laws} \labelsect{pseudo-atomic-fixed-points}

\Theorem{distrib-fixed-iter} and \Theorem{distrib-finite-iter} both assume $d$ satisfies \refprop{d-sync-nil} and \refprop{d-sync-seq}.
A general property of $d$ that implies both \refprop{d-sync-nil} and \refprop{d-sync-seq} 
is that $d$ is a \emph{pseudo-atomic-fixed point}, that is, 
\begin{align}
  d = \Nil \nondet \Patx \Seq d \labeldef{pseudo-atomic-fixed-point}
\end{align}
for some pseudo-atomic command $\Patx$.
Pseudo-atomic fixed points satisfy the following basic properties.

\begin{lemmax}[pafp-basic-properties]
If $\Ata$ is an atomic command and,
for some pseudo-atomic command $\Patx$, 
$d$ is a pseudo-atomic fixed point such that,
$d = \Nil \nondet \Patx \Seq d$, then,
\begin{align}
  d \sync \cgd{p} & = \cgd{p} & \mbox{for any set of states $p$} \labelprop{pafp-sync-test} \\
  d \sync \Ata\Seq c &= (\Patx \sync \Ata) \Seq (d \sync c) \labelprop{pafp-sync-atomic-seq} \\
  d \sync \Ata & = \Patx \sync \Ata \labelprop{pafp-sync-atomic} \\
  d \sync \Ata \Seq c & = (d \sync \Ata) \Seq (d \sync c) \labelprop{pafp-distrib-atomic-seq}
\end{align}
\end{lemmax}

\begin{proof}
For \refprop{pafp-sync-test},
the proof follows by expanding $d$ using \refdef{pseudo-atomic-fixed-point},
and then simplifying using \refprop{sync-distrib-nondet}, \refax{test-sync-test}, \refprop{pseudo-sync-nil}, \refax{whole-if-first},
and the fact that $\Magic$ is the identity of non-deterministic choice:
$d \sync \cgd{p}
= (\Nil \nondet \Patx \Seq d) \sync \cgd{p}
= (\Nil \sync \cgd{p}) \nondet (\Patx \Seq d \sync \cgd{p})
= \cgd{p} \nondet \Magic
= \cgd{p}
$.

To verify \refprop{pafp-sync-atomic-seq} we reason as follows.
\begin{align*}&
  d \sync \Ata \Seq c 
 \Equals*[expanding $d$ using \refdef{pseudo-atomic-fixed-point} and distributing using \refprop{sync-distrib-nondet}]
  (\Nil \sync \Ata \Seq c) \nondet (\Patx \Seq d \sync \Ata \Seq c)
\Equals*[applying \refprop{pseudo-sync-nil} and \refax{whole-if-first} to show $\Nil \sync \Ata \Seq c = \Magic$]
  \Patx \Seq d \sync \Ata \Seq c
 \Equals*[using \refprop{pseudo-sync-pseudo} noting that $\Ata$ is atomic and hence also pseudo-atomic]
  (\Patx \sync \Ata) \Seq (d \sync c)
\end{align*}

Property \refprop{pafp-sync-atomic} follows by 
using the fact that $\Nil$ the unit of sequential composition,
applying \refprop{pafp-sync-atomic-seq},
and then \refprop{pafp-sync-test}: 
$d \sync \Ata 
= d \sync \Ata \Seq \Nil
= (\Patx \sync \Ata) \Seq (d \sync \Nil)
= \Patx \sync \Ata$.

Property \refprop{pafp-distrib-atomic-seq} then follows by applying \refprop{pafp-sync-atomic-seq} and \refprop{pafp-sync-atomic}. 
\end{proof}

The proof that a pseudo-atomic fixed point distributes over a sequential composition \refprop{d-sync-seq},
that is, $d \sync c_1 \Seq c_2 = (d \sync c_1) \Seq (d \sync c_2)$,
makes use of the fact that the command $c_1$ can be decomposed into its finite behaviours of length $k$ for all $k \in \nat$, 
plus its infinite behaviours.
The finite behaviours of length $k$ are given by, $\SyncAtomicId^k \sync c_1$,
recalling that $\SyncAtomicId$ is the atomic identity of $\sync$
and hence $\SyncAtomicId^k$ allows exactly $k$ transitions.
The infinite behaviours of $c_1$ correspond to $\Inf{\SyncAtomicId} \sync c_1$.
We first show the distributive law holds if $c_1$ is restricted to its finite traces of length $k$.

\begin{lemmax}[pseudo-atomic-distrib-length]
For pseudo-atomic fixed point $d$ and $k \in \nat$,
\begin{align}
  (d \sync \SyncAtomicId^k \sync c_1) \Seq (d \sync c_2) = d \sync (\SyncAtomicId^k \sync c_1) \Seq c_2 . \labelprop{pseudo-atomic-distrib-length}
\end{align}
\end{lemmax}

\begin{proof}
The proof is by induction on $k$ using the properties of pseudo-atomic fixed points.
We use the expanded form for $c_1$, that is, $c_1 = \Pre{p_n} \Seq (\cgd{p_t} \nondet \Nondet_{(\Ata,c') \in C} (\Ata \Seq c'))$,
for some sets of states $p_n$ and $p_t$, and set of pairs of of atomic command and commands $C$.
For $k = 0$.
\begin{align*}&
  (d \sync \SyncAtomicId^0 \sync c_1) \Seq (d \sync c_2)
 \Equals*[as $\SyncAtomicId^0 = \Nil$ by \refax{iter-zero} and $d \sync \Nil = \Nil$ by \refprop{pafp-sync-test}]
  (\Nil \sync c_1) \Seq (d \sync c_2)
 \Equals*[by \refprop{expanded-nil} using the expanded form of $c_1$]
  \Pre{p_n} \Seq \cgd{p_t} \Seq (d \sync c_2)
 \Equals*[distributing by \refprop{test-command-sync-command}
   (as $d \sync \Magic = \Magic$ by \refprop{pafp-sync-test})
   and \refprop{assert-command-sync-command}]
 d \sync (\Pre{p_n} \Seq \cgd{p_t} \Seq c_2)
 \Equals*[by \refprop{expanded-nil} and $\Nil = \SyncAtomicId^0$ by \refax{iter-zero}]
 d \sync ((\SyncAtomicId^0 \sync c_1) \Seq c_2)
 \end{align*}
For the inductive case, we assume \refprop{pseudo-atomic-distrib-length} for $k$ and show it holds for $k+1$.
\begin{align*}&
  (d \sync \SyncAtomicId^{k+1} \sync c_1) \Seq (d \sync c_2) 
 \Equals*[by \refprop{expanded-succ} using the expanded form for $c_1$ and $\SyncAtomicId \sync \Ata = \Ata$ by \refax{sync-atomic-id}]
  \left(d \sync
    \left(\Pre{p_n} \Seq \Nondet_{(\Ata,c') \in C} (\Ata \Seq (\SyncAtomicId^k \sync c'))\right)
  \right)
  \Seq (d \sync c_2)
 \Equals*[distributing assertion by \refprop{assert-command-sync-command},
          then distributing by \refprop{sync-distrib-Nondet} and \refax{seq-distrib-Nondet-right}]
  \Pre{p_n} \Seq \left(\Nondet_{(\Ata,c') \in C} (d \sync (\Ata \Seq (\SyncAtomicId^k \sync c'))) \Seq (d \sync c_2) \right)
\end{align*}
\begin{align*}
 \Equals*[distributing $d$ by \refprop{pafp-distrib-atomic-seq}]
  \Pre{p_n} \Seq \left(\Nondet_{(\Ata,c') \in C} (d \sync \Ata) \Seq (d \sync (\SyncAtomicId^k \sync c')) \Seq (d \sync c_2) \right)
 \Equals*[using the induction hypothesis]
  \Pre{p_n} \Seq \left(\Nondet_{(\Ata,c') \in C} (d \sync \Ata) \Seq (d \sync ((\SyncAtomicId^k \sync c') \Seq  c_2)) \right)
 \Equals*[distributing $d$ by \refprop{pafp-distrib-atomic-seq}]
  \Pre{p_n} \Seq \left(\Nondet_{(\Ata,c') \in C} d \sync (\Ata \Seq (\SyncAtomicId^k \sync c') \Seq  c_2) \right)
 \Equals*[distributing by \refax{seq-distrib-Nondet-right} and then \refprop{sync-distrib-Nondet}, and then \refprop{assert-command-sync-command}]
  d \sync \left( \left( \Pre{p_n} \Seq \Nondet_{(\Ata,c') \in C}  \Ata \Seq (\SyncAtomicId^k \sync c') \right) \Seq c_2 \right)
 \Equals*[by \refprop{expanded-succ} using the expanded form for $c_1$ and $\SyncAtomicId \sync \Ata = \Ata$ by \refax{sync-atomic-id}] 
  d \sync (\SyncAtomicId^{k+1} \sync c_1) \Seq c_2
 \qedhere
\end{align*}
\end{proof}

\reflem{pseudo-atomic-distrib-length} can then be generalised to the case where $c$ is restricted any finite number of steps.

\begin{lemmax}[pseudo-atomic-distrib-fin]
For pseudo-atomic fixed point $d$,
\begin{align}
  (d \sync \Fin{\SyncAtomicId} \sync c_1) \Seq (d \sync c_2) = d \sync ((\Fin{\SyncAtomicId} \sync c_1) \Seq c_2) .
\labelprop{pseudo-atomic-distrib-fin}
\end{align}
\end{lemmax}

\begin{proof}
We have
\begin{align*}&
  (d \sync \Fin{\SyncAtomicId} \sync c_1) \Seq (d \sync c_2)
\Equals*[using \refdef{iter-finite} and distributing using \refax{sync-distrib-Nondet} and \refax{seq-distrib-Nondet-right} ]
  \Nondet_{k \in \nat} (d \sync \SyncAtomicId^k \sync c_1) \Seq (d \sync c_2) 
 \Equals*[by \reflem{pseudo-atomic-distrib-length}]
   \Nondet_{k \in \nat} d \sync ((\SyncAtomicId^k \sync c_1) \Seq c_2)
 \Equals*[distributing using \refax{sync-distrib-Nondet} and \refax{seq-distrib-Nondet-right} and then \refax{sync-distrib-Nondet} a second time]
  d \sync ((\Nondet_{k \in \nat} \SyncAtomicId^k) \sync c_1) \Seq c_2
 \Equals*[using \refdef{iter-finite}]
  d \sync (\Fin{\SyncAtomicId} \sync c_1) \Seq c_2
 \qedhere
\end{align*}
\end{proof}

For the case where $c$ is restricted to take an infinite number of
steps, our distribution property hold trivially because $\Inf{\SyncAtomicId}$ left annihilates.

\begin{lemmax}[inf-annihilates]
$(\Inf{c_0} \sync c_1) \Seq c_2 = \Inf{c_0} \sync c_1$
\end{lemmax}

\begin{proof}
From the fact that $\Inf{c_0}$ left annihilates, $\Inf{c_0} = \Inf{c_0} \Seq \Magic$
and as $\Magic$ is a degenerate test, we can apply \refprop{final-test}, and the fact that $\Magic$ is a left annihilator, 
and then reverse the steps:
$(\Inf{c_0} \sync c_1) \Seq c_2 
= (\Inf{c_0} \Seq \Magic \sync c_1) \Seq c_2 
= (\Inf{c_0} \sync c_1) \Seq \Magic
= \Inf{c_0} \Seq \Magic \sync c_1
= \Inf{c_0} \sync c_1
$.
\end{proof}

\begin{lemmax}[pseudo-atomic-distrib-inf]
For pseudo-atomic fixed point $d$,
\begin{align}
  (d \sync \Inf{\SyncAtomicId} \sync c_1) \Seq (d \sync c_2) = d \sync ((\Inf{\SyncAtomicId} \sync c_1) \Seq c_2) .
\labelprop{pseudo-atomic-distrib-inf}
\end{align}
\end{lemmax}

\begin{proof}
Applying \reflem{inf-annihilates} twice, we get:
$ (d \sync \Inf{\SyncAtomicId} \sync c_1) \Seq (d \sync c_2)
= d \sync (\Inf{\SyncAtomicId} \sync c_1)
= d \sync ((\Inf{\SyncAtomicId} \sync c_1) \Seq c_2)
$.
\end{proof}

The command $\Om{\SyncAtomicId}$ is the identity of $\sync$ \refax{sync-id}, and so
from properties \refprop{isolation} and  \refprop{sync-distrib-nondet}  we have for any command $c$,
\begin{align}\labelprop{decomp}
  c = \Om{\SyncAtomicId} \sync c 
  = (\Fin{\SyncAtomicId} \nondet \Inf{\SyncAtomicId}) \sync c
  = (\Fin{\SyncAtomicId}\sync c) \nondet (\Inf{\SyncAtomicId} \sync c).
\end{align}
and so distribution of pseudo-atomic-fixed points over sequential
composition follow from \reflem{pseudo-atomic-distrib-fin} and
\reflem{pseudo-atomic-distrib-inf}.

\begin{theoremx}[pseudo-atomic-distrib-seq]
If $d$ is a pseudo-atomic fixed point, for any commands $c_1$ and $c_2$,
\(
  d \sync c_1 \Seq c_2 = (d \sync c_1) \Seq (d \sync c_2) .
\)
\end{theoremx}

\begin{proof}
\begin{align*}&
  (d \sync c_1) \Seq (d \sync c_2)
 \Equals*[using decomposition \refprop{decomp} and distributing using \refprop{seq-distrib-nondet-right}]
  (d \sync \Fin{\SyncAtomicId}\sync c_1) \Seq (d \sync c_2)
  \nondet
  (d \sync \Inf{\SyncAtomicId}\sync c_1) \Seq (d \sync c_2)
 \Equals*[by \reflem{pseudo-atomic-distrib-fin} and \reflem*{pseudo-atomic-distrib-inf}]
  d \sync ((\Fin{\SyncAtomicId}\sync c_1) \Seq  c_2)
  \nondet
  d \sync ((\Inf{\SyncAtomicId}\sync c_1) \Seq  c_2)
 \Equals*[distributing using \refprop{sync-distrib-nondet} and \refprop{seq-distrib-nondet-right}]  
  d \sync ((\Fin{\SyncAtomicId}\sync c_1 \nondet \Inf{\SyncAtomicId}\sync c_1) \Seq  c_2)
 \Equals*[using decomposition \refprop{decomp}]
  d \sync (c_1 \Seq  c_2) 
 \qedhere
\end{align*}
\end{proof}

Corollaries of \Theorem{pseudo-atomic-distrib-seq} and \refprop{pafp-sync-test} 
using \Theorem{distrib-fixed-iter} and \Theorem{distrib-finite-iter} are the following.
\begin{corx}[pseudo-atomic-distrib-fixed-iter]
If $d$ is a pseudo-atomic fixed point then, for all $i \in \nat$,
\(
  d \sync c^i = (d \sync c)^i .
\)
\end{corx}

\begin{corx}[pseudo-atomic-distrib-finite-iter]
If $d$ is a pseudo-atomic fixed point then,
\(
  d \sync \Fin{c} = \Fin{(d \sync c)} .
\)
\end{corx}

Two pseudo-atomic fixed points synchronise to give a pseudo-atomic fixed point.
\begin{lemmax}[pseudo-atomic-conjunction]
If both $d_1$ and $d_2$ are pseudo-atomic fixed points, so is $d_1 \sync d_2$.
\end{lemmax}

\begin{proof}
Because $d_1$ and $d_2$ are pseudo-atomic fixed points, for some pseudo-atomic commands $\Patx_1$ and $\Patx_2$, 
$d_1 = \Nil \nondet \Patx_1 \Seq d_1$ and
$d_2 = \Nil \nondet \Patx_2 \Seq d_2$.
Hence using \refax{test-sync-test}, \refax{whole-if-first}, \refprop{pseudo-sync-nil} and \refprop{pseudo-sync-pseudo},
$d_1 \sync d_2 
=  (\Nil \nondet \Patx_1 \Seq d_1) \sync (\Nil \nondet \Patx_2 \Seq d_2)
= \Nil \nondet (\Patx_1 \Seq d_1 \sync \Patx_2 \Seq d_2)
= \Nil \nondet (\Patx_1 \sync \Patx_2) \Seq (d_1 \sync d_2)
$,
and it is straightforward to show $\Patx_1 \sync \Patx_2$ is a pseudo-atomic command 
by expanding $\Patx_1$ to $\Ata_1 \nondet \Atb_1 \Seq \Abort$ and $\Patx_2$ to $\Ata_2 \nondet \Atb_2 \Seq \Abort$, 
for some atomic commands $\Ata_1$, $\Atb_1$, $\Ata_2$ and $\Atb_2$, and simplifyling.
\end{proof}

\section{Application to rely/guarantee concurrency} \labelsect{rely-guar}

To support rely/guarantee concurrency we follow the approach of Aczel \cite{Aczel83} and
record both the transitions made by a thread $T$ (called program or $\pstepd$ transitions)
and transitions made by the environment of $T$ (called environment or $\estepd$ transitions)
in the traces of $T$.
The environment transitions record transitions made by any thread running in parallel with $T$.
We make use of two atomic commands 
that represent transitions by the thread, $\cpstep{r}$, and transitions by the environment of $T$, $\cestep{r}$,
for $r$ a relation between program states. More precisely,
\begin{description}
\item[$\cpstep{r}$] is an atomic command that can perform a single program ($\pstepd$) transition from state $\sigma$ to state $\sigma'$, if $(\sigma,\sigma') \in r$,
and then terminate, and
\item[$\cestep{r}$] is an atomic command that can perform a single environment ($\estepd$) transition from $\sigma$ to $\sigma'$, if $(\sigma,\sigma') \in r$, 
and then terminate.
\end{description}
The universal relation between states is denoted $\universalrel$.
The atomic command $\cpstepd$ can perform any single program transition \refdef{cpstepd},
the atomic command $\cestepd$ can perform any single environment transition \refdef{cestepd},
the atomic command $\cstep{r}$ can perform any single transition (program or environment) satisfying $r$,
and
the atomic command $\cstepd$ can perform any single transition \refdef{cstepd}.
Note the bold fonts on the left for $\cpstepd$, $\cestepd$ and $\cstepd$.
\begin{eqncolumns}
  \cpstepd & \defs &\cpstep{\universalrel} \labeldef{cpstepd} \\
  \cestepd & \defs &\cestep{\universalrel} \labeldef{cestepd} 
\secondcolumn
  \cstep{r} & \defs & \cpstep{r} \nondet \cestep{r} \labeldef{cstep} \\
  \cstepd & \defs & \cstep{\universalrel} \labeldef{cstepd}
\end{eqncolumns}

\paragraph{Distributing guarantees and relies.}

A program guarantee condition $g$, where $g$ is a relation between states, 
requires that all program transitions from $\sigma$ to $\sigma'$ are in $g$, i.e.\ $(\sigma,\sigma') \in g$.
It places no constrains on environment transitions.
A program guarantee can be encoded as a command, $\guar{g}$, 
that is defined as an iteration of the atomic command $\cpstep{g} \nondet \cestepd$ \refdef{guar}.
\begin{align}
  \guar{g} & \defs \Om{(\cpstep{g} \nondet \cestepd)} \labeldef{guar}
\end{align}
A rely condition $r$ corresponds to an assumption that all environment transitions satisfy $r$.
It is encoded as the command, $\rely{r}$, that allows any transitions but aborts if the environment makes a transition not satisfying $r$.
It is defined as an iteration of the pseudo-atomic command $\cstepd \nondet \cestep{\overline{r}} \Seq \Abort$,
where $\overline{r}$ is the complement of the relation $r$  \refdef{rely}.
The rely command has an alternative equivalent form \refprop{rely-alt}
because $\cstepd = \cpstepd \nondet \cestep{r} \nondet \cestep{\overline{r}}$ and 
$\cestep{\overline{r}} \Seq \Abort \refsto \cestep{\overline{r}}$,
so that $\cestep{\overline{r}} \nondet \cestep{\overline{r}} \Seq \Abort = \cestep{\overline{r}} \Seq \Abort$.
\begin{eqncolumns}
  \rely{r} & \defs \Om{(\cstepd \nondet \cestep{\overline{r}} \Seq \Abort)} \labeldef{rely} 
\secondcolumn
  \rely{r} & = \Om{(\cpstepd \nondet \cestep{r} \nondet \cestep{\overline{r}} \Seq \Abort)} \labelprop{rely-alt}
\end{eqncolumns}
The use of abort to represent failure of a rely command is similar to 
the use of abort in the sequential refinement calculus to represent failure of an assertion command, $\Pre{p}$.
Both commands encode \emph{assumptions}: 
the assertion command encodes the assumption that $p$ holds initially, 
and the rely command encodes the assumption that environment transitions satisfy $r$.

Both $\guar{g}$ and $\rely{r}$ are possibly infinite iterations of pseudo-atomic commands 
and hence the distributive laws detailed in the previous sections apply for distributing 
them with a synchronisation operator over a sequential composition (giving \refprop{guar-sync-distrib-seq} and \refprop{rely-sync-distrib-seq}) 
and hence parallel (giving \refprop{guar-par-distrib-seq} and \refprop{rely-par-distrib-seq}) and 
weak conjunction (giving \refprop{guar-conj-distrib-seq} and \refprop{rely-conj-distrib-seq}),
and hence they distribute via synchronisation (and hence parallel or weak conjunction) into a fixed or finite iteration.

\paragraph{Other pseudo-atomic fixed points.}

The command $\Term$ is the most general command 
that performs only a finite number of program transitions but does not constrain environment transitions \refdef{term},
and the command $\Fair$ is the most general command that disallows an infinite contiguous sequence of environment transitions \refdef{fair}.
\begin{eqncolumns}
  \Term & \defs \Fin{\cstepd} \Seq \Om{\cestepd} \labeldef{term} 
\secondcolumn
  \Fair & \defs \Fin{\cestepd} \Seq \Om{(\cpstepd \Seq \Fin{\cestepd})} \labeldef{fair}
\end{eqncolumns}
While neither of these commands is expressed as an iteration of a (single) pseudo-atomic command,
both $\Term$ and $\Fair$ are pseudo-atomic fixed points 
because they satisfy the following fixed point equations.
\begin{eqncolumns}
  \Term & = \Nil \nondet \cstepd \Seq \Term
\secondcolumn
  \Fair & = \Nil \nondet \cstepd \Seq \Fair
\end{eqncolumns}
Hence both $\Term$ and $\Fair$ distribute over sequential composition 
and into fixed and finite iterations.

The command $\Idle$ allows a finite number of program transitions that do not change the program state
(i.e.\ they satisfy the identity relation on states, $\Id$) but does not constrain environment transitions.
\begin{align}
  \Idle & \defs \guar{\Id} \together \Term \labeldef{idle}
\end{align}
Because both $\guar{\Id}$ and $\Term$ are pseudo-atomic fixed points, by \reflem{pseudo-atomic-conjunction} so is $\Idle$,
and hence it distributes over sequential compositions and into fixed and finite iterations.

\paragraph{Evolution invariants.}

The command, $\evolve{r}$, represents an evolution invariant \cite{ColletteJones00a},
which is useful when every transition of a command, both program and environment, is expected to satisfy a relation $r$.
It guarantees program transitions satisfy $r$ and relies on environment transitions satisfying $r$ \refdef{evolve}.
It may be written in the form of an iteration of the pseudo-atomic command, 
$\cpstep{r} \nondet \cestep{r} \nondet \cestep{\overline{r}} \Seq \Abort$ \refprop{evolve-alt}.
\begin{eqncolumns}
  \evolve{r} & \defs \guar{r} \together \rely{r} \labeldef{evolve}
\secondcolumn
  \evolve{r} & = \Om{(\cpstep{r} \nondet \cestep{r} \nondet \cestep{\overline{r}} \Seq \Abort)} \labelprop{evolve-alt}
\end{eqncolumns}
Evolution invariants are pseudo-atomic fixed points and hence they distribute over sequential composition and into fixed and finite iterations.

\paragraph{Generalised invariants.}\labelsect{invariants}

A generalised invariant is a property that, if it holds initially, holds in every state of an execution trace \cite{ReynoldsCraft81}.
The invariant command, $\inv{p}$, is defined in terms of an evolution invariant 
with a relation $\postr{p}$ that ensures $p$ holds in the final state for all possible initial states \refdef{postr}.
The evolution invariant is prefixed by an assumption, $\Pre{p}$, that the initial state is in $p$ \refdef{inv}.
\begin{eqncolumns}
  \prer{p} & \defs \{ (\sigma, \sigma') \spot \sigma \in p \} \labeldef{prer} \\
  \postr{p} & \defs \{ (\sigma, \sigma') \spot \sigma' \in p \} \labeldef{postr}
\secondcolumn
  \inv{p} & \defs \Pre{p} \Seq \evolve{\postr{p}} \labeldef{inv}
\end{eqncolumns}
It can be shown that,
\begin{align}
  \Pre{p} \Seq \evolve{\postr{p}} = \Pre{p} \Seq \evolve{(\overline{\prer{p}} \union \postr{p})} = \Pre{p} \Seq \evolve{(\prer{p} \inter \postr{p})}
\end{align}
where $\overline{\prer{p}} \union \postr{p}$ can be read as if $p$ holds before, it holds after.
Any of these three alternatives can then be used for reasoning about invariants.
In fact, any of the alternatives could be used as the definition of a generalised invariant;
we chose \refdef{inv} because it makes reasoning in the theorems below simpler.

A generalised invariant command, $\inv{p}$, is not a pseudo-atomic fixed point (due to the assertion at its beginning)
but it is defined in terms of $\evolve{\postr{p}} $, which is a pseudo-atomic fixed point
and hence in proving distributive laws for $\inv{p}$, 
one can make use of the distributive laws for evolution invariants.

As possibly infinite iteration, $\Om{c}$, of a command $c$ is the greatest fixed point of $(\lambda y \spot \Nil \nondet c \Seq y)$,
it satisfies the following standard unfolding \refprop{omega-unfold} and fixed point induction \refprop{omega-induct} laws.
\begin{align}
  \Om{c} & = \Nil \nondet c \Seq \Om{c}  \labelprop{omega-unfold} \\
  \Om{c} \Seq d & \refsto x & \mbox{if } d \nondet c \Seq x \refsto x \labelprop{omega-induct}
\end{align}
If a command, $c$, maintains a test $\cgd{p}$, then so does $\Om{c}$.

\begin{lemmax}[maintain-inv]
If $\cgd{p} \Seq c \Seq \cgd{p} = \cgd{p} \Seq c$ then, 
$\Pre{p} \Seq \Om{c} \Seq \cgd{p} = \Pre{p} \Seq \Om{c}$.
\end{lemmax}

\begin{proof}
Refinement from right to left holds trivially as one can always introduce a test because $\Nil \refsto \cgd{p}$, for any test $\cgd{p}$.
For the refinement from left to right we first apply the Galois connection between tests and assertions \refprop{galois-test-assert},
so that we are required to show: 
$\Om{c} \Seq \cgd{p} \refsto \cgd{p} \Seq \Pre{p} \Seq \Om{c}$,
or as the test absorbs the following assertion \refprop{test-assert},
$\Om{c} \Seq \cgd{p} \refsto \cgd{p} \Seq \Om{c}$,
which holds by iteration induction \refprop{omega-induct} if, 
$\cgd{p} \nondet c \Seq \cgd{p} \Seq \Om{c} \refsto \cgd{p} \Seq \Om{c}$,
which we show by introducing a second occurrence of $\cgd{p}$,
using the assumption, 
distributing by \refax{seq-distrib-Nondet-left}, 
and then folding the iteration \refprop{omega-unfold}:
$\cgd{p} \nondet c \Seq \cgd{p} \Seq \Om{c}
\refsto \cgd{p} \nondet  \cgd{p} \Seq c \Seq  \cgd{p}\Seq \Om{c}
=  \cgd{p} \nondet \cgd{p} \Seq c \Seq \Om{c}
= \cgd{p} \Seq (\Nil \nondet c \Seq \Om{c})
= \cgd{p} \Seq \Om{c}
$.
\end{proof}
\noindent
That lemma can be applied to show that a generalised invariant, $\inv{p}$, establishes $p$.
\begin{lemmax}[inv-establish]
For any set of states $p$,~ $\inv{p} \Seq \cgd{p} = \inv{p}$.
\end{lemmax}

\begin{proof}
The definition of a generalised invariant \refdef{inv} expands $\inv{p}$ to $\Pre{p} \Seq \evolve{\postr{p}}$,
and by \refprop{evolve-alt}, 
$\Pre{p} \Seq \evolve{\postr{p}} = \Pre{p} \Seq \Om{(\cpstep{\postr{p}} \nondet \cestep{\postr{p}} \nondet \cestep{\overline{\postr{p}}} \Seq \Abort)}$
and hence we can apply \reflem{maintain-inv} with $c$ as $(\cpstep{\postr{p}} \nondet \cestep{\postr{p}} \nondet \cestep{\overline{\postr{p}}} \Seq \Abort)$.
The proviso for \reflem{maintain-inv} requires us to show, 
\[
  \cgd{p} \Seq (\cpstep{\postr{p}} \nondet \cestep{\postr{p}} \nondet \cestep{\overline{\postr{p}}} \Seq \Abort) \Seq \cgd{p} = 
    \cgd{p} \Seq (\cpstep{\postr{p}} \nondet \cestep{\postr{p}} \nondet \cestep{\overline{\postr{p}}} \Seq \Abort) ,
\]
which holds because the commands $\cpstep{\postr{p}}$ and $\cestep{\postr{p}}$ both establish $p$
(i.e.\ $\cpstep{\postr{p}} \Seq \cgd{p} = \cpstep{\postr{p}}$ and $\cestep{\postr{p}} \Seq \cgd{p} = \cestep{\postr{p}}$)
and $\Abort$ is an annihilator from the left so that,
$\cestep{\overline{\postr{p}}} \Seq \Abort \Seq \cgd{p} = \cestep{\overline{\postr{p}}} \Seq \Abort$.
\end{proof}

\begin{lemmax}[inv-sync-establish]
$\inv{p} \sync c = (\inv{p} \sync c) \Seq \Pre{p}$.
\end{lemmax}

\begin{proof}
The proof uses \reflem{inv-establish} then \refprop{final-test} and \refprop{test-assert} then reverses \refprop{final-test} and \reflem{inv-establish}:
$\inv{p} \sync c 
= \inv{p} \Seq \cgd{p} \sync c
= (\inv{p} \sync c) \Seq \cgd{p} \Seq \Pre{p}
= (\inv{p} \Seq \cgd{p} \sync c) \Seq \Pre{p}
= (\inv{p} \sync c) \Seq \Pre{p}
$.
\end{proof}

The above lemma can then be used to show that a generalised invariant distributes over a sequential composition.
\begin{theoremx}[inv-distrib-sync-seq]
$\inv{p} \sync c_1 \Seq c_2 = (\inv{p} \sync c_1) \Seq (\inv{p} \sync c_2)$
\end{theoremx}

\begin{proof}
The proof makes use of the fact that $\evolve{}$ is a pseudo-atomic fixed point
because it can be expressed as the iteration of a pseudo-atomic command \refprop{evolve-alt}.
\begin{align*}&
  \inv{p} \sync c_1 \Seq c_2
 \Equals*[definition of $\inv{}$ \refdef{inv} and pull out assertion using \refprop{assert-command-sync-command}]
  \Pre{p} \Seq (\evolve{\postr{p}} \sync c_1 \Seq c_2)
 \Equals*[by \Theorem{pseudo-atomic-distrib-seq}]
  \Pre{p} \Seq (\evolve{\postr{p}} \sync c_1) \Seq (\evolve{\postr{p}} \sync c_2)
 \Equals*[move assertion in by \refprop{assert-command-sync-command}; rewrite as an invariant \refdef{inv}; apply \reflem*{inv-sync-establish}]
  (\inv{p} \sync c_1) \Seq \Pre{p} \Seq (\evolve{\postr{p}} \sync c_2)
 \Equals*[push assertion in using \refprop{assert-command-sync-command} and then rewrite as an invariant \refdef{inv}]
  (\inv{p} \sync c_1) \Seq (\inv{p} \sync c_2)
 \qedhere
\end{align*}
\end{proof}

An invariant distributes into a fixed iteration 
but requires an initial assertion, $\Pre{p}$, on the right in the following lemma to cope with the zero iterations case.
\begin{lemmax}[inv-distrib-fixed-iter]
$\inv{p} \sync c^i = \Pre{p} \Seq (\inv{p} \sync c)^i$
\end{lemmax}

\begin{proof}
The proof is by induction on $i$.
For $i=0$, 
$\inv{p} \sync c^0 
= \Pre{p} \Seq (\evolve{\postr{p}} \sync \Nil)
= \Pre{p} \Seq \Nil
= \Pre{p} \Seq (\inv{p} \sync c)^0
$.
We assume for $i$ and show it holds for $i+1$.
\begin{align*}&
  \inv{p} \sync c^{i+1} 
 \Equals*[by \refax{iter-succ} and \Theorem{inv-distrib-sync-seq}]
  (\inv{p} \sync c) \Seq (\inv{p} \sync c^i)
 \Equals*[inductive hypothesis]
  (\inv{p} \sync c) \Seq \Pre{p} \Seq (\inv{p} \sync c)^i
 \Equals*[by \reflem{inv-sync-establish}]
  (\inv{p} \sync c) \Seq (\inv{p} \sync c)^i
 \Equals*[as $\inv{p} = \Pre{p} \Seq \inv{p}$ from \refdef{inv}; pull out $\Pre{p}$ using \refprop{assert-command-sync-command}]
  \Pre{p} \Seq (\inv{p} \sync c) \Seq (\inv{p} \sync c)^i
 \Equals*[by \refax{iter-succ}]
  \Pre{p} \Seq (\inv{p} \sync c)^{i+1}
 \qedhere
\end{align*}
\end{proof}

Distribution over a fixed iteration can then be lifted to distribution over a finite iteration.
\begin{lemmax}[inv-distrib-finite-iter]
$\inv{p} \sync \Fin{c} = \Pre{p} \Seq \Fin{(\inv{p} \sync c)}$
\end{lemmax}

\begin{proof}
The proof uses the decomposition of a finite iteration as a choice over all fixed iterations \refdef{iter-finite},
applies \reflem{inv-distrib-fixed-iter} to each,
distributes the assertion out of the choice \refax{seq-distrib-Nondet-left}, and recomposes the iteration \refdef{iter-finite}:
$\inv{p} \sync \Fin{c} 
= \inv{p} \sync \Nondet_{i \in \nat} c^i
= \Nondet_{i \in \nat} (\inv{p} \sync c^i)
= \Nondet_{i \in \nat} \Pre{p} \Seq (\inv{p} \sync c)^i
= \Pre{p} \Seq \Nondet_{i \in \nat} (\inv{p} \sync c)^i
= \Pre{p} \Seq \Fin{(\inv{p} \sync c)}
$.
\end{proof}

\section{Conclusions}

Distributive laws are important in algebraic reasoning in arithmetic and logic,
and they are equally important for reasoning about programs \cite{HoareHayesEtcFull87}.
In Concurrent Kleene Algebra \cite{DBLP:journals/jlp/HoareMSW11}
and our earlier theory \cite{AFfGRGRACP,FM2016atomicSteps,FMJournalAtomicSteps},
laws for distributing over sequential composition and into iterations were only refinements in a single direction.
The current paper strengthens the distributive laws to be equalities.
Such strengthening requires the ability to decompose a command into an expanded form \refax{expanded-form}
in terms of its immediately terminating behaviours, its immediately aborting behaviours, and
its behaviours consisting of an atomic transition followed by a continuation command.
In addition, the ability to partition the reasoning about a command into its behaviours of a given length is crucial.
The proofs of the strengthened distributive laws are somewhat more complex than the corresponding weak versions;
they require the full power of the synchronous algebra.

The current paper focuses on distributing the abstract synchronisation operator, $\sync$, over sequential compositions 
and into fixed and finite iterations. 
As $\sync$ is an abstraction of both parallel composition ($\parallel$) and weak conjunction ($\together$),
we immediately gain the distributive laws for both these operators.

The strong distributive laws are shown to hold provided the command, $d$, 
being distributed is a pseudo-atomic fixed point.
This is a sufficient condition but not a necessary condition, 
for example, a generalised invariant command, $\inv{p}$, is not a pseudo-atomic fixed point
but it does distribute over sequential compositions and into fixed and finite iterations.
However, the proof of the laws for a generalised invariant 
makes use of the fact that it is defined in terms of an evolution invariant, $\evolve{\postr{p}}$,
which is a pseudo-atomic fixed point, 
and hence the proofs of the laws for generalised invariants depend on those for pseudo-atomic fixed points.

While CKA \cite{DBLP:journals/jlp/HoareMSW11} and Synchronous Kleene Algebra \cite{Pris10} only consider partial correctness,
our concurrent refinement algebra handles total correctness,
and hence includes infinite iteration as well as finite iteration.
Further these approaches only provide weak (i.e.\ refinement) distributive laws,
whereas our approach provides strong (i.e.\ equality) laws.

Our generalised invariant command was inspired by that of Morgan and Vickers \cite{TaIitRC-SCP},
who introduced an invariant command in the sequential refinement calculus, with essentially the same semantics as here,
that is, the invariant $p$ is assumed initially and maintained automatically within the scope of the command.
The main difference is that our generalised invariant command needs to deal with fine-grained concurrency.
Their command has the syntax $|\![\inv{p} \spot c]\!|$, 
and is defined in terms of an extended weakest precondition semantics that takes an invariant (a predicate) as an additional parameter,
whereas our invariant command is defined in terms of our language primitives 
and combined with a command using weak conjunction, an operator that is not available in their theory.

An important application of the strengthened distributive laws is for performing data refinements,
where an abstract representation of a data structure is replaced by a lower-level, more efficient implementation state.
The invariant command \refdef{inv} can be used to encode a coupling invariant for a data refinement:
it relates the abstract state of a data type to the concrete implementation state.
As part of proving a data refinement a (coupling) invariant is distributed into construct being data refined, 
then the construct refined in that context, 
and finally distributing the invariant is reversed; 
this requires the distributive laws to be applicable in both directions.

\paragraph{Future work.}

The iteration $\Om{\Patx}$, in which $\Patx$ is a pseudo-atomic command, is a pseudo-atomic fixed point,
in fact, the greatest fixed point.
We believe that $\Om{\Patx}$ also satisfies both the following laws.
\begin{align}
  \Om{\Patx} \sync \Inf{c} & = \Inf{(\Om{\Patx} \sync c)} \labelprop{sync-distrib-infinite-iter} \\
  \Om{\Patx} \sync \Om{c} & = \Om{(\Om{\Patx} \sync c)} \labelprop{sync-distrib-iter1}
\end{align}
Note that these do not hold in general with $\Om{\Patx}$ replaced by $\Fin{\Patx}$
because with $\Fin{\Patx}$, the left sides only allow a finite number of transitions because $\Fin{\Patx}$ does,
whereas the right sides allow an infinite number of transitions 
because $\Om{(\Fin{\Patx} \sync c)}$ allows an infinite number of iterations of $\Fin{\Patx} \sync c$.
Rely and guarantee commands are both defined in the form $\Om{\Patx}$ for $\Patx$ a pseudo-atomic command
and hence they both satisfy \refprop{sync-distrib-infinite-iter} and \refprop{sync-distrib-iter1}.

Similar laws can be developed for distributing $\together$ over $\parallel$,
\begin{align}
  d \together (c_1 \parallel c_2) = ( d \together c_1) \parallel (d \together c_2) .
\end{align}
We believe this distributive law holds if $d$ is restricted to an iteration of a pseudo-atomic command, $\Patx$, 
for which $\Patx = \Patx \parallel \Patx$.
Suitable commands for $d$ are thus $\guar{g}$ and $\evolve{r}$, but not $\rely{r}$.

\paragraph{Acknowledgements.}
Thanks are due to
Joakim von Wright for introducing us to program algebra,
and
Callum Bannister
and 
Dan Nathan,
for feedback on ideas presented in this paper
and/or 
contributions to the supporting Isabelle/HOL theories.
Special thanks go to Cliff Jones for his continual feedback and encouragement.
This work is supported by the 
Australian Research Council
under their Discovery Program Grant No.~DP190102142.

\bibliographystyle{plain}
\bibliography{ms}

\ifarxiv
\else
\printindex
\fi

\end{document}

%% file: rely-guar.tex
\begin{picture}(0,0)%
\includegraphics{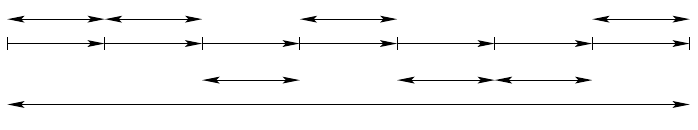}%
\end{picture}%
\setlength{\unitlength}{2565sp}%
\begingroup\makeatletter\ifx\SetFigFont\undefined%
\gdef\SetFigFont#1#2#3#4#5{%
  \reset@font\fontsize{#1}{#2pt}%
  \fontfamily{#3}\fontseries{#4}\fontshape{#5}%
  \selectfont}%
\fi\endgroup%
\begin{picture}(8580,1587)(1111,-3814)
\put(9676,-2836){\makebox(0,0)[lb]{\smash{{\SetFigFont{8}{9.6}{\rmdefault}{\mddefault}{\updefault}{\color[rgb]{0,0,0}$\checkmark$}%
}}}}
\put(1801,-2686){\makebox(0,0)[b]{\smash{{\SetFigFont{8}{9.6}{\rmdefault}{\mddefault}{\updefault}{\color[rgb]{0,0,0}$\estepd$}%
}}}}
\put(1801,-2386){\makebox(0,0)[b]{\smash{{\SetFigFont{8}{9.6}{\rmdefault}{\mddefault}{\updefault}{\color[rgb]{0,0,1}$r$}%
}}}}
\put(3001,-2686){\makebox(0,0)[b]{\smash{{\SetFigFont{8}{9.6}{\rmdefault}{\mddefault}{\updefault}{\color[rgb]{0,0,0}$\estepd$}%
}}}}
\put(4201,-2686){\makebox(0,0)[b]{\smash{{\SetFigFont{8}{9.6}{\rmdefault}{\mddefault}{\updefault}{\color[rgb]{0,0,0}$\pstepd$}%
}}}}
\put(5476,-2686){\makebox(0,0)[b]{\smash{{\SetFigFont{8}{9.6}{\rmdefault}{\mddefault}{\updefault}{\color[rgb]{0,0,0}$\estepd$}%
}}}}
\put(6601,-2686){\makebox(0,0)[b]{\smash{{\SetFigFont{8}{9.6}{\rmdefault}{\mddefault}{\updefault}{\color[rgb]{0,0,0}$\pstepd$}%
}}}}
\put(7801,-2686){\makebox(0,0)[b]{\smash{{\SetFigFont{8}{9.6}{\rmdefault}{\mddefault}{\updefault}{\color[rgb]{0,0,0}$\pstepd$}%
}}}}
\put(9001,-2686){\makebox(0,0)[b]{\smash{{\SetFigFont{8}{9.6}{\rmdefault}{\mddefault}{\updefault}{\color[rgb]{0,0,0}$\estepd$}%
}}}}
\put(3001,-2386){\makebox(0,0)[b]{\smash{{\SetFigFont{8}{9.6}{\rmdefault}{\mddefault}{\updefault}{\color[rgb]{0,0,1}$r$}%
}}}}
\put(5401,-2386){\makebox(0,0)[b]{\smash{{\SetFigFont{8}{9.6}{\rmdefault}{\mddefault}{\updefault}{\color[rgb]{0,0,1}$r$}%
}}}}
\put(9001,-2386){\makebox(0,0)[b]{\smash{{\SetFigFont{8}{9.6}{\rmdefault}{\mddefault}{\updefault}{\color[rgb]{0,0,1}$r$}%
}}}}
\put(1126,-2836){\makebox(0,0)[rb]{\smash{{\SetFigFont{8}{9.6}{\rmdefault}{\mddefault}{\updefault}{\color[rgb]{0,0,1}$p$}%
}}}}
\put(5476,-3736){\makebox(0,0)[b]{\smash{{\SetFigFont{8}{9.6}{\rmdefault}{\mddefault}{\updefault}{\color[rgb]{1,0,0}$q$}%
}}}}
\put(4201,-3436){\makebox(0,0)[b]{\smash{{\SetFigFont{8}{9.6}{\rmdefault}{\mddefault}{\updefault}{\color[rgb]{1,0,0}$g$}%
}}}}
\put(7801,-3436){\makebox(0,0)[b]{\smash{{\SetFigFont{8}{9.6}{\rmdefault}{\mddefault}{\updefault}{\color[rgb]{1,0,0}$g$}%
}}}}
\put(6601,-3436){\makebox(0,0)[b]{\smash{{\SetFigFont{8}{9.6}{\rmdefault}{\mddefault}{\updefault}{\color[rgb]{1,0,0}$g$}%
}}}}
\put(1201,-3061){\makebox(0,0)[b]{\smash{{\SetFigFont{8}{9.6}{\rmdefault}{\mddefault}{\updefault}{\color[rgb]{0,0,0}$\sigma_0$}%
}}}}
\put(2401,-3061){\makebox(0,0)[b]{\smash{{\SetFigFont{8}{9.6}{\rmdefault}{\mddefault}{\updefault}{\color[rgb]{0,0,0}$\sigma_1$}%
}}}}
\put(3601,-3061){\makebox(0,0)[b]{\smash{{\SetFigFont{8}{9.6}{\rmdefault}{\mddefault}{\updefault}{\color[rgb]{0,0,0}$\sigma_2$}%
}}}}
\put(4801,-3061){\makebox(0,0)[b]{\smash{{\SetFigFont{8}{9.6}{\rmdefault}{\mddefault}{\updefault}{\color[rgb]{0,0,0}$\sigma_3$}%
}}}}
\put(6001,-3061){\makebox(0,0)[b]{\smash{{\SetFigFont{8}{9.6}{\rmdefault}{\mddefault}{\updefault}{\color[rgb]{0,0,0}$\sigma_4$}%
}}}}
\put(7201,-3061){\makebox(0,0)[b]{\smash{{\SetFigFont{8}{9.6}{\rmdefault}{\mddefault}{\updefault}{\color[rgb]{0,0,0}$\sigma_5$}%
}}}}
\put(8401,-3061){\makebox(0,0)[b]{\smash{{\SetFigFont{8}{9.6}{\rmdefault}{\mddefault}{\updefault}{\color[rgb]{0,0,0}$\sigma_6$}%
}}}}
\put(9601,-3061){\makebox(0,0)[b]{\smash{{\SetFigFont{8}{9.6}{\rmdefault}{\mddefault}{\updefault}{\color[rgb]{0,0,0}$\sigma_7$}%
}}}}
\end{picture}%